\documentclass[a4paper]{llncs}

\usepackage[english]{babel}
\usepackage[utf8x]{inputenc}
\usepackage{xifthen} 

\usepackage{tablefootnote}
\usepackage{makecell}
\usepackage{xspace}
\usepackage{pgfplots}
\usepackage{pgfplotstable}
\usepackage{multirow}

\usepackage{hyperref}  

\usepackage{listings}
\usepackage{csvsimple} 

\usepackage[inline,shortlabels]{enumitem} 
\newlist{inlinelist}{enumerate*}{1}
\setlist*[inlinelist,1]{%
  label=(\roman*),
}

\usepackage[final,nomargin,inline,index]{fixme} 
\fxusetheme{color}
\FXRegisterAuthor{bart}{anbart}{\color{magenta} {\underline{bart}}}
\definecolor{liviogreen}{rgb}{0, 0.5, 0}
\FXRegisterAuthor{livio}{anlivio}{\color{liviogreen} {\underline{livio}}}
\FXRegisterAuthor{ste}{anste}{\color{red} {\underline{ste}}}

\usepackage{cleveref}

\usetikzlibrary{pgfplots.dateplot}
\usetikzlibrary{matrix,chains,positioning,decorations.pathreplacing}
\usetikzlibrary{patterns,calc,fit,arrows}
\usetikzlibrary{shapes.geometric}

\hypersetup{
  colorlinks   = true, 
  urlcolor     = blue, 
  linkcolor    = blue, 
  citecolor   = red 
}
\hypersetup{final} 

\definecolor{lightgray}{rgb}{0.83, 0.83, 0.83}
\definecolor{codeBG}{rgb}{.98,.98,.98}%
\definecolor{keyword}{HTML}{7F0055}

\lstdefinelanguage{bdl}{
	commentstyle=\color{lightgray},
	morecomment=[l]{//},
	morecomment=[s]{/*}{*/},
	morestring=[b]",
    stringstyle=\color{blue},
	classoffset=0,
	morekeywords={String,Int,Bool,Double,Timestamp,Iterable},
	keywordstyle=\color{keyword}\bfseries,
    classoffset=1,
    morekeywords={script,collection,object,mongo},
	keywordstyle=\color{blue}\bfseries,
    classoffset=2,
    morekeywords={all,last,in,if,then,else},
	keywordstyle=\color{ForestGreen}\bfseries
}

\lstdefinelanguage{scala}{
  morekeywords={%
          abstract,case,catch,class,def,do,else,extends,%
          false,final,finally,for,foreach,forSome,if,implicit,import,lazy,%
          match,new,null,object,override,package,private,protected,%
          return,sealed,super,this,throw,trait,true,try,type,%
          val,var,while,with,yield},
  keywordstyle=\color{blue}\bfseries,
  otherkeywords={=>,<-,<\%,<:,>:,\#,@},
  sensitive=true,
  morecomment=[l]{//},
  morecomment=[n]{/*}{*/},
  morestring=[b]",
  morestring=[b]',
  morestring=[b]"""
}[keywords,comments,strings]

\newcommand{\ifempty}[3]{%
  \ifthenelse{\isempty{#1}}{#2}{#3}%
}

\newcommand{\ifzero}[3]{%
  \ifthenelse{\equal{#1}{0}}{#2}{#3}%
}

\newcommand{\inputBDL}[1]{%
    \lstinputlisting[%
    firstline=1,
        language=scala,
        ]{#1}
}%

\newcommand{\inputBDLlinenos}[1]{%
    \lstinputlisting[%
    firstline=1,
        language=scala,
    numbers=left
        ]{#1}
}%

\newlength\replength
\newcommand\repfrac{.1}

\newcommand\rulewidth{.6pt}
\newcommand\tdashfill[1][\repfrac]{\cleaders\hbox to \replength{%
	\smash{\rule[\arraystretch\ht\strutbox]{\repfrac\replength}{\rulewidth}}}\hfill}

\newcommand\tdotfill[1][\repfrac]{\cleaders\hbox to \replength{%
	\smash{\raisebox{\arraystretch\dimexpr\ht\strutbox-.1ex\relax}{.}}}\hfill}

\newcommand{\seq}[1]{\langle {#1} \rangle}

\def\etc{\emph{etc}.\@\xspace}

\newcommand{\eg}{e.g.\@\xspace}
\newcommand{\ie}{i.e.\@\xspace}

\newcommand{\opreturn}{OP\_RETURN\xspace}

\newcommand{\codefont}{\fontsize{10}{10}\selectfont}
\newcommand{\code}[1]{{\tt {#1}}}

\def\transColor{\color{darkgray}}
\newcommand{\transFmt}[1]{{\transColor{\sf #1}}}

\newcommand{\transT}[1][]{\transFmt{t}_{\transColor{#1}}}

\newcommand{\transIn}[2][]{{\sf \transColor{in}}\ifempty{#1}{}{[{#1}]}: {#2}}
\newcommand{\transInScript}[2][]{{\sf \transColor{in-script}}\ifempty{#1}{}{[{#1}]}: {#2}}
\newcommand{\transOutScript}[3][]{{\sf \transColor{out-script}}\ifempty{#1}{}{[{#1}]}({#2}): {#3}}

\newcommand{\transValue}[2][]{{\sf \transColor{value}}\ifempty{#1}{}{[{#1}]}: {#2}}

\newcommand{\btc}{\textup{%
  \leavevmode
  \vtop{\offinterlineskip 
    \setbox0=\hbox{B}%
    \setbox2=\hbox to\wd0{\hfil\hskip-.03em
    \vrule height .3ex width .15ex\hskip .08em
    \vrule height .3ex width .15ex\hfil}
    \vbox{\copy2\box0}\box2}}}

\newcommand{\BTC}{\mbox{\textit{BTC}}\xspace}
\newcommand{\USD}{\mbox{\textit{USD}}\xspace}

\newcommand{\lineno}[1]{{\tt\codefont {\textcolor{magenta}{#1}}}}

\newtoggle{draft}

\newcommand{\mytitle}{A general framework for blockchain analytics}
\newcommand{\mygithub}{https://github.com/bitbart/blockchain-analytics-api}
\newcommand{\mylastblock}{473100}	
\newcommand{\mylastdate}{2017/06/27}

\makeatletter
\newcommand\footnoteref[1]{\protected@xdef\@thefnmark{\ref{#1}}\@footnotemark}
\makeatother

\begin{document}

\title{\mytitle}

\titlerunning{\mytitle} 
\author{Massimo Bartoletti\inst{1} \and Andrea Bracciali\inst{2}
\and Stefano Lande\inst{1} \and \\ Livio Pompianu\inst{1}}

\authorrunning{M. Bartoletti, A. Bracciali, S. Lande, L. Pompianu}
\tocauthor{Massimo Bartoletti, Andrea Bracciali, Stefano Lande, Livio Pompianu}
\institute{University of Cagliari, Italy \\
\email{\{bart,lande,livio.pompianu\}@unica.it}
\and
University of Stirling, UK \\
\email{abb@cs.stir.ac.uk}}
  
\maketitle
\setcounter{footnote}{0}

\begin{abstract}
  Modern cryptocurrencies exploit decentralised blockchains to record a public and unalterable history of transactions.
  Besides transactions, further information is stored for different, and often undisclosed, purposes, making the blockchains a rich and increasingly growing source of valuable information, in part of difficult interpretation. 
  Many data analytics have been developed, 
  mostly based on specifically designed and ad-hoc engineered approaches.
  We propose a general-purpose framework, seamlessly supporting data analytics on both Bitcoin and Ethereum --- currently the two most prominent cryptocurrencies. 
  Such a framework allows us to integrate relevant blockchain data with data from other sources, 
  and to organise them in a database, either SQL or NoSQL. 
  Our framework is released as an open-source Scala library.
  We illustrate the distinguishing features of our approach on a set of significant use cases, 
  which allow us to empirically compare ours to other competing proposals, 
  and evaluate the impact of the database choice on scalability.
\end{abstract}

\section{Introduction}
\label{sec:intro}

The last few years have witnessed a steady growth in interest on blockchains,
driven by the success of Bitcoin and, more recently, of Ethereum. 
This has fostered the research on several aspects of blockchain technologies, 
from their theoretical foundations 
--- both cryptographic~\cite{Bonneau15ieeesp,Garay15eurocrypt} and economic~\cite{Luu15csfw,Schrijvers16fc} ---
to their security and privacy~\cite{Androulaki13fc,Bonneau14mixcoin,Gervais16ccs,Karame15tissec,Meiklejohn16cacm}.

Among the research topics emerging from blockchain technologies,
one that has received major interest is the analysis of the data 
stored in blockchains.
Indeed, the two main blockchains contain several gigabytes of data
($\sim$130GB for Bitcoin, $\sim$300GB for Ethereum),
that only in part are related to currency transfers.
Developing analytics on these data allows us to obtain several insights,
as well as economic indicators that help to predict market trends.

Many works on data analytics have been recently published, 
addressing an\-onym\-ity issues,
\eg by de-anonymising users~\cite{Meiklejohn13imc,Meiklejohn16cacm,Ober13futureinternet,Reid13analysis},
clustering transactions~\cite{Harrigan16unreasonable,Spagnuolo14bitiodine},
or evaluating anonymising services~\cite{Moser17anonymous}.
Other analyses have addressed criminal activities,
\eg by studying
denial-of-service attacks~\cite{baqer2016stressing,Vasek14fc},
ransomware~\cite{Liao16ecrime},
and various financial frauds \cite{Moser13ecrime,Moser14bitcoin,Vasek15fc}.
Many statistics on Bitcoin and Ethereum exist,
measuring \eg economic indicators~\cite{lischke2016analyzing,ron2013quantitative},
transaction fees~\cite{Moser15fc},
the usage of metadata~\cite{BP17bitcoin},
\etc

A common trait of these works is that
they create \emph{views} of the blockchain which 
contain all the data needed for the goals of the analysis.
In many cases, this requires to
combine data \emph{within} the blockchain
with data from the \emph{outside}.
These data are retrieved from a variety of sources,
\eg blockchain explorers, wikis, discussion forums, and dedicated sites
(see~\Cref{tab:related} for a brief survey).
Despite such studies share several common operations, 
\eg, scanning all the blocks and the transactions in the blockchain, 
converting the value of a transaction from bitcoins to \USD, \etc,
researchers so far tended to implement ad-hoc tools for their analyses, 
rather than reusing standard libraries.
Further, most of the few available tools have limitations, 
\eg they feature a fixed set of analytics,
or they do not allow to combine blockchain data with external data,
or they are not amenable to be updated.
The consequence is that the same functionalities have been 
implemented again and again as new analytics have been developed,
as witnessed by~\Cref{tab:related}.

In this context, we believe that the introduction of an efficient, 
modular and general-purpose abstraction layer to manage internal and external information 
is key for blockchain data analytics, along the lines of the software engineering best practices of \emph{reuse}.

  \begin{table}[t]
  \centering
  \resizebox{\columnwidth}{!}{%
    \begin{tabular}{ |l|l|l| } 
      \hline
      \textbf{Analysis goal} & \textbf{Gathered data} & \textbf{Sources} \\ 
      \hline 
      \hline
      Anonymity & 
                  \makecell[l]{
                  Transactions graph\\ 
      \opreturn metadata\\
      IP addresses\\
      address tags\\
      address tags} &
                      \makecell[l]{
                      bitcoind~\cite{Meiklejohn13imc,Meiklejohn16cacm,Moser17anonymous,Reid13analysis,Spagnuolo14bitiodine}, forum.bitcoin.org~\cite{Reid13analysis}\\
      bitcoind~\cite{Moser17anonymous}\\
      bitcoin faucet~\cite{Reid13analysis}, blockchain.info~\cite{Moser17anonymous}\\
      blockchain.info~\cite{Meiklejohn13imc,Meiklejohn16cacm,Spagnuolo14bitiodine}, bitcointalk.org~\cite{Meiklejohn13imc,Meiklejohn16cacm,Spagnuolo14bitiodine}\\
      bitcoin-otc.com~\cite{Spagnuolo14bitiodine}, bitfunder.org~\cite{Spagnuolo14bitiodine}}
      \\
      \hline
      \begin{tabular}{l} Market \\ analytics \end{tabular} &
                         \makecell[l]{
                         Transactions graph\\
      IP addresses\\
      address tags\\
      trade data} &
                    \makecell[l]{
                    bitcoind~\cite{lischke2016analyzing}, blockexplorer.com~\cite{ron2013quantitative}\\
      blockchain.info, ipinfo.io~\cite{lischke2016analyzing}\\
      blockchain.info~\cite{lischke2016analyzing}\\
      bitcoincharts.com~\cite{lischke2016analyzing}}
      \\
      \hline
      Cyber-crime & 
                    \makecell[l]{
                    Transactions graph\\
      mempool\\
      unconfirmed transactions\\
      no longer online services\\
      list of DDoS attacks\\ 
      mining pools\\
      trades on assets/services\\
      list of fraudulent services \\
      address tags\\
      exchange rate} &
                       \makecell[l]{
                       bitcoind~\cite{baqer2016stressing,Vasek15fc,Vasek14fc}, blockchain.info~\cite{Liao16ecrime,Moser13ecrime}, Bitcore~\cite{Bistarelli17ares} \\
      bitcoind~\cite{baqer2016stressing}\\
      bitcoind~\cite{baqer2016stressing}\\
      archive.org~\cite{Vasek15fc,Vasek14fc}\\
      bitcointalk.org~\cite{Vasek14fc}\\
      blockchain.info, \href{https://en.bitcoin.it/wiki/Category:Pool_Operators}{bitcoin wiki}~\cite{Vasek14fc}\\
      \href{https://en.bitcoin.it/wiki/Trade}{bitcoin wiki}~\cite{Vasek14fc}\\
      bitcointalk.org~\cite{Liao16ecrime,Vasek15fc}, badbitcoin.org~\cite{Vasek15fc}, cryptohyips.com~\cite{Vasek15fc}\\
      blockchain.info~\cite{Vasek15fc}\\
      bitcoincharts.com~\cite{Liao16ecrime,Vasek15fc}, quandl.com~\cite{Liao16ecrime}}
      \\
      \hline
      Metadata & 
                 \makecell[l]{
                 \opreturn transactions\\ 
      \opreturn identifiers} & 
                               \makecell[l]{
                               bitcoind~\cite{BP17bitcoin}\\
      kaiko.com, opreturn.org, \href{https://en.bitcoin.it/w/index.php?title=OP_RETURN\&oldid=61694}{bitcoin wiki}~\cite{BP17bitcoin}}
      \\
      \hline
      \begin{tabular}{l} Transaction \\ fees \end{tabular} &
                         \makecell[l]{
                         Transactions graph\\
      exchange rate\\
      mining pools} &
                      \makecell[l]{
                      bitcoind~\cite{Moser15fc}\\
      coindesk.com~\cite{Moser15fc} \\
      blockchain.info~\cite{Moser15fc}}
      \\
      \hline
    \end{tabular}%
  } %
  \caption{Data gathered by various blockchain analyses.}
  \label{tab:related}
  \vspace{-20pt}
\end{table}
  \paragraph{Contributions.}

The main contribution of this paper is 
a framework to create general-purpose analytics
on the block\-chains of Bitcoin and Ethereum.
The design of our tool is based on an exhaustive survey of the literature
on the analysis of blockchains.
The results of our survey, summarized in~\Cref{tab:related},
highlight the need to process external data besides those already 
present on the blockchain.
To this purpose, the workflow supported by our tool consists of two steps: 
\begin{inlinelist}
\item we construct a \emph{view} of the blockchain, 
also containing the needed external data,
and we save it in a database;
\item we analyse the view by using the query language of the DBMS.
\end{inlinelist}
The first step is supported by a new Scala library.
Distinguishably, we allow views to
be organised either as a MySQL database, or a MongoDB collection.
Our library supports the most commonly used external data,
\eg exchange rates, address tags, protocol identifiers,
and can be easily extended by linking the relevant data sources.
We evaluate the effectiveness of our framework by means of a set 
of paradigmatic use cases, which we distribute,
together with the source code of our library, 
under an open source license%
\footnote{\label{footnoteurl}\url{\mygithub}}.
We exploit our use cases to evaluate the performance of
SQL \emph{vs.} NoSQL databases 
for storing and querying blockchain views.
As a byproduct of our study, we provide a 
qualitative comparison of the other tools
for general-purpose blockchain analytics.
  \paragraph{Structure of the paper.}

In~\Cref{sec:api} we will illustrate our framework through a series of use cases.
We will perform experiments (using consumer hardware) which analyse 
blockchain metadata,
exchange rates
transactions fees, and
address tags.
In~\Cref{sec:evaluation} we will discuss some implementation details 
of our framework, and we will evaluate its effectiveness,
and the choice between SQL or NoSQL.
In~\Cref{sec:comparison} we will compare the existing 
general-purpose blockchain parsers with ours,
and finally in~\Cref{sec:conclusions} we will draw some conclusions.

\section{Background on Bitcoin}
\label{sec:bitcoin}

Bitcoin is a decentralized cryptocurrency~\cite{bitcoin,Bonneau15ieeesp}, 
that has recently reached a market capitalization of 100 \USD billions%
\footnote{
  Source: crypto-currency market capitalizations
  {\href{http://coinmarketcap.com}{\code{http://coinmarketcap.com}}}
}.
Bitcoin can be seen as a huge ledger of \emph{transactions},
which represent transfers of bitcoins ($\BTC$).
This ledger --- usually called \emph{blockchain} --- is 
maintained by a peer-to-peer network of nodes, 
and a consensus protocol ensures that it can only be updated consistently
(\eg, one cannot tamper with or remove an already-published transaction).

To give the intuition on how Bitcoin works,
we consider two transactions $\transT[0]$ and $\transT[1]$,
which we graphically represent as follows:%

\begin{center}
  \footnotesize
  \begin{tabular}{|l|}
    \hline
    \multicolumn{1}{|c|}{$\transT[0]$} \\
    \hline
    \transIn{$\cdots$} \\
    \transInScript{$\cdots$} \\
    \hline
    \transOutScript{$x$}{$F_0$} \qquad \\
    \transValue{$v_0$} \\
    \hline
  \end{tabular}
  \qquad\qquad
  \begin{tabular}{|l|}
    \hline
    \multicolumn{1}{|c|}{$\transT[1]$} \\
    \hline
    \transIn{$\mathit{hash}(\transT[0])$} \\
    \transInScript{$\sigma_1$} \\
    \hline
    \transOutScript{$\cdots$}{$\cdots$} \\
    \transValue{$v_1$} \\
    \hline
  \end{tabular}
\end{center}

The transaction $\transT[0]$ contains $v_0$ bitcoins,
which can be \emph{redeemed} by putting
on the blockchain a transaction (\eg, $\transT[1]$),
whose {\sf in} field is the cryptographic hash of the whole $\transT[0]$.
To redeem $\transT[0]$, the {\sf in-script} of $\transT[1]$
must contain a value $\sigma_1$ which makes the {\sf out-script} of $\transT[0]$
evaluate to true. 
In its general form, the {\sf out-script}
is a program in a (not Turing-complete) scripting language, 
featuring a limited set of logic, arithmetic, and cryptographic operators.
Typically, the {\sf out-script} is just a signature verification.

Now, assume that the blockchain contains $\transT [0]$,
not yet redeemed,
when someone tries to append~$\transT[1]$.
To validate this operation, the nodes of the Bitcoin network check that $v_1 \leq v_0$, 
and then they evaluate the {\sf out-script} $F_0$,
by instantiating its formal parameter $x$ to the value $\sigma_1$.
If, after the substitution, $F_0$ evaluates to true, 
then $\transT[1]$ redeems $\transT[0]$,
meaning that the value of $\transT[0]$ is transferred to the new transaction $\transT[1]$,
and $\transT[0]$ becomes no longer redeemable.
A new transaction can now redeem $\transT[1]$ by
satisfying its {\sf out-script}.

Bitcoin transactions may be more general than the ones illustrated by the previous example,
in that there can be multiple inputs and outputs.
Each output has an associated {\sf out-script} and value, 
and can be redeemed independently from others.
Consequently, {\sf in} fields must specify which output they are redeeming.
Similarly, a transaction with multiple inputs associates an {\sf in-script} to each of them.
To be valid, the sum of the values of all the inputs must be greater or equal to 
the sum of the values of all outputs.

The Bitcoin network is populated by a large set nodes, called \emph{miners}, 
which collect transactions from clients, 
and are in charge of appending the valid ones to the blockchain.
To this purpose, each miner keeps a local copy of the blockchain, 
and a set of unconfirmed transactions received by clients,
which it groups into \emph{blocks}.
The goal of miners is to add these blocks to the blockchain, 
in order to get a revenue.
Appending a new block $B_{i}$ to the blockchain
requires miners to solve a cryptographic puzzle,
which involves 
the hash $h(B_{i-1})$ of block $B_{i-1}$, 
a sequence of unconfirmed transactions $\seq{T_i}_i$, 
and some salt $R$. 
The goal of miners is to win the ``lottery'' for publishing the next block, 
\ie to solve the cryptopuzzle before the others;
when this happens, the miner receives a reward 
in newly generated bitcoins, 
and a \emph{fee} for each transaction included in the mined block
(the fee of a transaction is the difference between the values of its inputs and outputs).
If a miner claims the solution of the current cryptopuzzle, 
the others discard their attempts, 
update their local copies of the blockchain with the new block $B_i$, 
and start mining a new block on top of $B_i$.
In addition, miners are asked to verify the validity of 
the transactions in $B_i$ by executing the associated scripts.

\section{Creating blockchain analytics}
\label{sec:api}

We illustrate our framework through some case studies,
which, for uniformity, have been developed for the Bitcoin case.
We refer to our github repository\footnoteref{footnoteurl}
for some Ethereum examples.
Our library APIs provide the following Scala classes to represent 
the primitive entities of the blockchain:
\begin{itemize}

\item \code{BlockchainLib}: main library class. 
It provides the \code{getBlockchain} method, 
to iterate over \code{Block} objects.

\item \code{Block}: contains a list of transactions, 
and some block-related attributes (\eg, block hash and creation time).

\item \code{Transaction}: 
contains various related attributes (\eg, transaction hash and size).

\end{itemize}

The library constructs the above-mentioned Scala objects
by scanning a local copy of the blockchain.
It uses the client, either \href{https://bitcoin.org/en/bitcoin-core/}{Bitcoin Core} 
or \href{https://parity.io/}{Parity}, to have a direct access to the blocks, 
exploiting the provided indices.
For Bitcoin, it uses the \href{https://bitcoinj.github.io/}{BitcoinJ} library 
as a basis to represent the various kinds of objects,
while for Ethereum it uses suitable Scala representations.
The APIs allow constructed objects to be exported as MongoDB documents or MySQL records. 
In \href{https://www.mongodb.com}{MongoDB}
(a widespread non-relational DBMS)
a database is a set of \emph{collections}, each of them containing \emph{documents}.
Documents are lists of pairs \code{(k,v)}, 
where \code{k} is a string (called \emph{field name}), 
and \code{v} is either a value or a MongoDB document.
Conversely, \href{https://www.mysql.com/}{MySQL} implements the relational model, and represents
an objects as a record in a table.
In~\Cref{sec:api:myblockchain,sec:api:opreturn,sec:api:exchange,sec:api:fees,sec:api:tags} 
we develop a series of analytics on Bitcoin.
Full Scala code which builds the needed blockchain views, queries,
and analysis results can be found in the GitHub repository of the project\footnoteref{footnoteurl}.
  \subsection{A basic view of the Bitcoin blockchain}
\label{sec:api:myblockchain}

\begin{figure}[t]
	\centering
	\resizebox{\columnwidth}{!}{%
		\begin{minipage}{\columnwidth}
			\inputBDLlinenos{sources/myblockchain_BTC_MongoDB.txt}
		\end{minipage}
	}
	\vspace{-10pt}
	\caption{A basic view of the blockchain.}
	\label{lst:api:myblockchain}
\end{figure}

Since all the analyses shown in~\Cref{tab:related} explore the transaction
graph (\eg they investigate output values, timestamps, metadata, \etc),
our first case study focusses on a basic view of the Bitcoin blockchain
containing no external data.
The documents in the resulting collection represent transactions,
and they include:
\begin{inlinelist}
	\item the transaction hash;
	\item the hash of the enclosing block;
	\item the date in which the block was appended to the blockchain;
	\item the list of transaction inputs and outputs.
\end{inlinelist}

We show in~\Cref{lst:api:myblockchain} 
how to use our APIs to construct this collection. 
Lines~\lineno{1-2} are standard Scala instructions to define the \code{main} function.
The object \code{blockchain} constructed at line~\lineno{4} 
is a handle to the Bitcoin blockchain.
At line~\lineno{5} we setup the connection to Bitcoin Core, 
by providing the needed parameters
(user, password, and port),
and by indicating that we want to use the main network 
(alternatively, the parameter \code{TestNet} allows to use the test network).
At line~\lineno{6} we setup the connection to MongoDB
(alternatively, the parameter \code{MySQL} allows to use MySQL).
Since lines \lineno{1-6} are similar for all our case studies, 
for the sake of brevity we will omit them in the subsequent listings. 
We declare the target collection \code{myBlockchain} at line~\lineno{7}.
At this point, we start navigating the blockchain (from the origin block up 
to block number \mylastblock) to populate the collection.
To do that we iterate over the blocks (line~\lineno{9})
(note that \code{b => \{\dots\}} is an anonymous function, where \code{b} is a parameter, and \code{\{\dots\}} is its body),
and for each block we iterate over its transactions (at line~\lineno{10}).
For each transaction we append a new document to \code{myBlockchain} (lines~\lineno{11-16}).
This document is a set of fields of the form \code{(k,v)}, 
where \code{k} is the field name, and \code{v} is the associated value.
For instance, at line~\lineno{12} we stipulate that the field \code{txHash}
will contain the hash of the transaction, represented by \code{tx.hash}.
This value is obtained by the API \code{BitcoinTransaction}.

\begin{figure}[h]
	\centering
	\scriptsize
	\resizebox{\columnwidth}{!}{%
		\iftoggle{draft}{
			\includegraphics{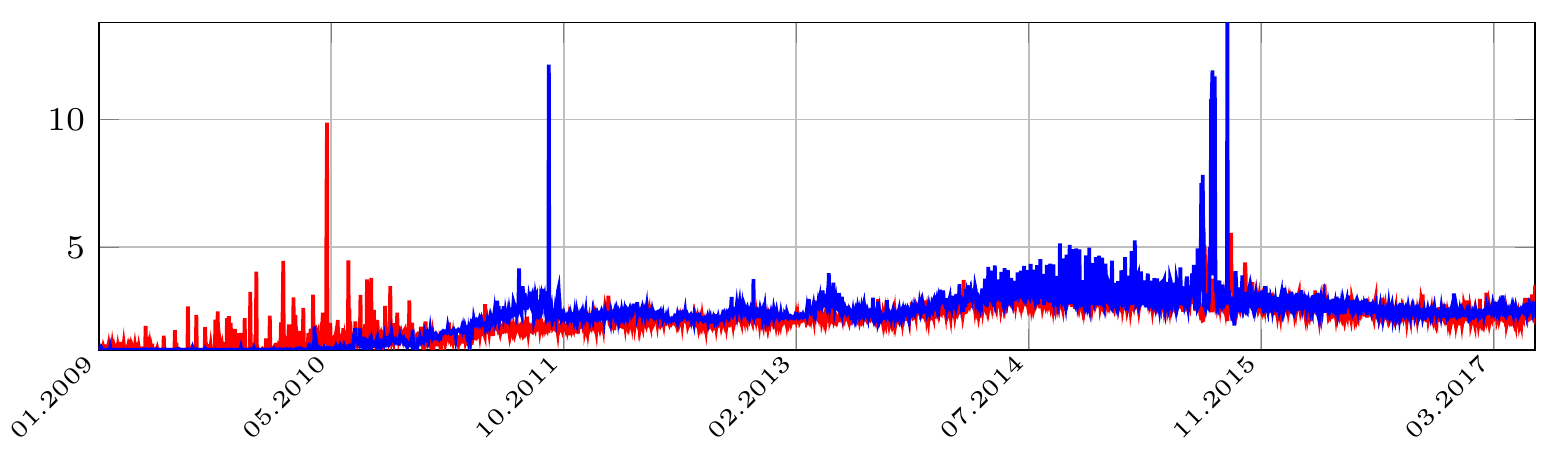}
		}
		{
			\begin{tikzpicture}
			\pgfplotsset{x coord trafo/.code={\pgfmathparse{#1}\pgfmathresult}}
			\begin{axis}[
			width  = 1\linewidth,
			height = 4cm,
			date coordinates in=x,
			xmajorgrids = true,
			ymajorgrids = true,
			xlabel absolute, xlabel style={yshift=-0.5cm},
			ylabel absolute, ylabel style={yshift=0.2cm},
			xticklabel=\tiny\month.\year,
			xticklabel style = {rotate=45,anchor=east},
			enlargelimits = false,
			scaled y ticks = false,
			extra y tick style={grid=major, tick label style={xshift=-1cm}},
			]
			\pgfplotstableread[col sep=comma]{queries/myBlockchain_BTC_MongoDB_Result1.csv}\data
			\addplot [color=red!100!white, thick, mark=none] table[x index = {0}, y index = {1}]{\data};
			\addplot [color=blue!100!white, thick, mark=none] table[x index = {0}, y index = {2}]{\data};
			\end{axis}
			\end{tikzpicture}
		}
	}
	\vspace{-15pt}
	\caption{Average number of inputs (red line) and outputs (blue line) by date.}
	\label{fig:pgf:myblockchain}
\end{figure}

Running this piece of code results in a view,
which we can process to obtain several standard statistics,
like \eg
the \href{https://blockchain.info/charts/n-transactions}{number of daily transactions}, 
their \href{https://bitinfocharts.com/comparison/bitcoin-transactionvalue.html}{average value},
the \href{https://blockchain.info/largest-recent-transactions}{largest recent transactions}, \etc
\footnote{Note that one could also perform these queries
	during the construction of the view.
	However, this would not be convenient in general, since 
	--- as we will see also in the following case studies ---
	many relevant queries can be performed on the same view.}
Hereafter we consider another kind of analysis,
\ie the evolution over the years of the number of transaction inputs and outputs.
To this purpose, we run the MongoDB query shown in~\Cref{lst:mongo:myblockchain}.
The query first groups the documents with the same date.
Then, for each group, it computes the average number of inputs and outputs. 
Finally, the results are sorted in ascending order. 
The results of the query are graphically rendered 
in~\Cref{fig:pgf:myblockchain},
which shows the average number of inputs/outputs by date.
We see that, after an initial phase,
the average number of inputs and outputs has stabilised between 2 and 3.
This is mainly due to the fact that most transactions are published through standard wallets,
which try to minimise the number of inputs;
a typical transaction has two outputs, one to perform the payment and the other for the change. 
We also observe a few peaks in the number of inputs and outputs, 
which are probably related to experimentation of new services, 
like \eg \href{https://en.bitcoin.it/wiki/CoinJoin}{CoinJoin}.

\begin{figure}[t]
	\centering
	\resizebox{\columnwidth}{!}{%
		\begin{minipage}{\columnwidth}
			\inputBDL{queries/myBlockchain_BTC_MongoDB_Query1.txt}
		\end{minipage}
	}
	\vspace{-10pt}
	\caption{A query to estimate the average number of inputs and outputs by date.}
	\label{lst:mongo:myblockchain}
\end{figure}

  \subsection{Analysing \opreturn metadata}
\label{sec:api:opreturn}

\begin{figure}[t]
	\centering
	\resizebox{\columnwidth}{!}{%
		\begin{minipage}{\columnwidth}
			\inputBDLlinenos{sources/opreturn_BTC_MongoDB.txt}
		\end{minipage} 
	}
	\vspace{-10pt}
	\caption{Exposing \opreturn metadata.}
	\label{lst:api:opreturn}
\end{figure}

Besides being used as a cryptocurrency,
Bitcoin allows for appending a few bytes of metadata to transaction outputs.
This is done preeminently through the \href{https://en.bitcoin.it/wiki/OP_RETURN}{\opreturn operator}.
Several protocols exploit this feature to implement blockchain-based applications,
like \eg digital assets and notarization services~\cite{BP17bitcoin}.

We now construct a view of the blockchain 
which exposes the protocol metadata.
More specifically, the entries of the view represent transaction outputs, and are composed of: 
\begin{inlinelist}
	\item the hash of the transaction containing the output;
	\item the date in which the transaction has been appended to the blockchain;
	\item the name of the protocol that produced the transaction;
	\item the metadata contained in the \opreturn script.
\end{inlinelist}
\Cref{lst:api:opreturn} shows the Scala code to construct this collection
(we omit the declaration of the \code{main} method, already shown in~\Cref{lst:api:myblockchain}). 
At line~\lineno{3} we scan the blockchain, starting from block 290,000 since
\opreturn transactions were only relayed as standard transactions after the \href{https://bitcoin.org/en/release/v0.9.0}{release 0.9.0 of Bitcoin Core}.
We then iterate through transactions at line~\lineno{4}, 
and through their outputs at line~\lineno{5}.
We append a new document to our collection (lines~\lineno{7-11})
whenever the output of the corresponding transaction is an \opreturn (line~\lineno{6}).
The method \code{OpReturn.getApplication} of our APIs
takes as input a piece of metadata, and returns the name of the associated protocol.
This is inferred by the results of the analysis in~\cite{BP17bitcoin}.

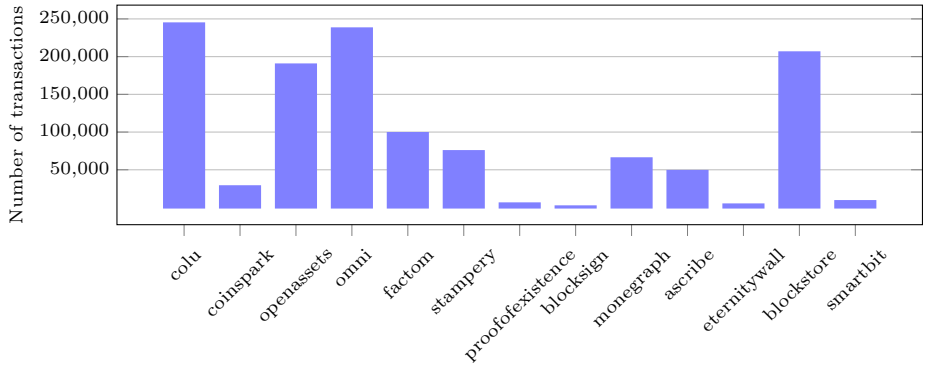
\begin{figure}[h]
  \centering
  \scriptsize
  \resizebox{\columnwidth}{!}{%
    \begin{tikzpicture}
       \begin{axis}[
        width = 1\linewidth,
        symbolic x coords={colu,coinspark,openassets,omni,factom,stampery,proofofexistence,blocksign,monegraph,ascribe,eternitywall,blockstore,smartbit},
        height = 4.5cm,
        xtick=data,
        ybar,
        bar width=15pt,
        xlabel absolute, 
        ylabel absolute, ylabel style={yshift=0.1cm},
        tick label style={/pgf/number format/fixed},
        ytick={50000,100000,150000,200000,250000},
        enlargelimits = true,
        ymajorgrids = true,
        xtick pos=left,
        xticklabel style={rotate=45, font=\scriptsize},
        ylabel={Number of transactions},
        scaled y ticks = false,
        ]
        \pgfplotstableread[col sep=comma]{queries/opreturn_BTC_MongoDB_Result1.csv}\data
        \addplot [fill=blue!50,draw=blue!50,solid,thick,mark=none] table[x index = {0}, y index = {1}]{\data};
      \end{axis}
    \end{tikzpicture}
  }
  \vspace{-15pt}
  \caption{Number of transactions per protocol (only protocols with $>1000$ transactions).}
  \label{fig:pgf:opreturn}
\end{figure}

The obtained view can be used to perform various analyses.
For instance, we show in \Cref{fig:pgf:opreturn} the number of transactions associated with 
the most used protocols (only those with at least 1000 transactions).
The protocol with the highest number of transactions is \href{https://www.colu.com/}{Colu}, 
which is used to certify and transfer the ownership of physical assets. 
The second most used protocol is \href{http://www.omnilayer.org/}{Omni}, 
followed by \href{https://github.com/blockstack/blockchain-id/wiki/Blockstore}{Blockstore}, 
a key-value store upon which other protocols are based.

  \subsection{Exchange rates}
\label{sec:api:exchange}

\begin{figure}[t]
	\centering
	\resizebox{\columnwidth}{!}{%
		\begin{minipage}{\columnwidth}
			\inputBDLlinenos{sources/exchange_BTC_MongoDB.txt}
		\end{minipage}
	}
	\vspace{-10pt}
	\caption{Exposing exchange rates.}
	\label{lst:api:exchange}
\end{figure}

Several analyses in~\Cref{tab:related} use exchange rates 
for quantifying the economic impact of various phenomena 
(\eg cyber-crime attacks, transaction fees, business activities).  
In this~\namecref{sec:api:exchange} 
we analyse how the value transferred in transactions
is affected by the exchange rate between \USD and \BTC
over the years.
Since exchange rates are not stored in the Bitcoin blockchain,
we need to obtain these data from an external source,
\eg the \href{http://www.coindesk.com/price/bitcoin-price-index/}{Coindesk APIs}.
Using these data, we construct a blockchain view where
each transaction is associated with the exchange rate 
at the time it has been appended to the blockchain.
More specifically, we construct a MongoDB collection %
whose documents represent transactions containing: 
\begin{inlinelist}
\item the transaction hash;
\item the date in which the transaction has been appended to the blockchain;
\item the sum of its output values (in \BTC);
\item the exchange rate between \BTC and \USD in such date.
\end{inlinelist}

\Cref{lst:api:exchange} shows the Scala code which builds this collection, using our APIs.
At line~\lineno{1} we declare the collection that we are going to build, \code{txWithRates}.
At lines~\lineno{3-4} we iterate over all the transactions in the Bitcoin blockchain.
For each one, at lines~\lineno{5-9} we add a new document to \code{txWithRates}. 
The total amount of \BTC sent by the current transaction is stored in the field 
\code{outputSum} (line~\lineno{8}). 
The exchange rate is obtained by invoking the method \code{Exchange} of our APIs (line~\lineno{9}).
This method takes a date and retrieves from Coindesk the exchange rate \BTC/\USD in that date.

\begin{figure}[h]
  \centering
  \scriptsize
  \resizebox{\columnwidth}{!}{%
    \begin{tikzpicture}
      \begin{axis}[
        width = 1\linewidth,
        symbolic x coords = {0-300,300-600,600-900,900-1200,1200-1500,1500-1800,1800-2100},
        height = 4.5cm,
        xtick=data,
        ybar,
        bar width=15pt,
        xlabel absolute, xlabel style={yshift=-0.5cm},
        ylabel absolute, ylabel style={yshift=0cm},
        enlargelimits = true,
        ymajorgrids = true,
        ylabel={\BTC},
        ]
        \pgfplotstableread[col sep=comma]{queries/exchange_BTC_MongoDB_Result1.csv}\data
        \addplot [fill=blue!50,draw=blue!50,solid,thick,mark=none] table[x index = {0}, y index = {1}]{\data};
      \end{axis}
    \end{tikzpicture}
  }
  \vspace{-15pt}
  \caption{Average value of outputs (in \BTC) by exchange rate.}
  \label{fig:pgf:exchange}
\end{figure}
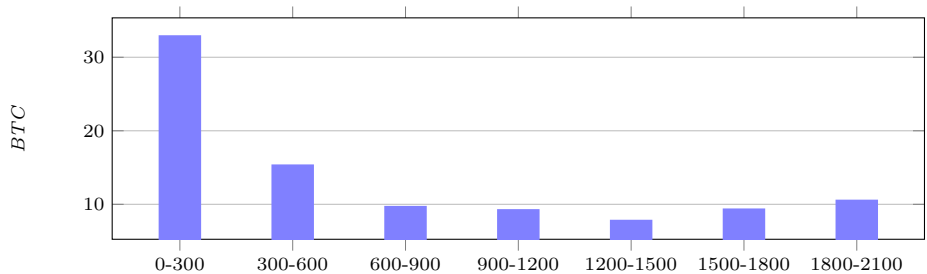

We can analyse the obtained collection in many ways, 
in order to study how exchange rates are related to the movements of currency in Bitcoin.
For instance, one can obtain statistics about
the \href{https://blockchain.info/charts/estimated-transaction-volume-usd}{daily transaction volume} in \USD,
the \href{http://blockchain.info/charts/market-cap}{market capitalization},
the \href{https://bitinfocharts.com/top-100-richest-bitcoin-addresses.html}{list of richest addresses}, \etc
Hereafter, we measure the average value of outputs (in \BTC) 
of the transactions in intervals of exchange rates.
The diagram in~\Cref{fig:pgf:exchange} shows the results of this analysis,
where we have split exchange rates in 7 intervals of equal size.
In the first five intervals we observe the expected behaviour, 
\ie the value of outputs decreases as the exchange rate increases.
Perhaps surprisingly, the last two intervals show an increase in the value of outputs 
when the value \BTC has surpassed 1500 \USD.
This may be explained by speculative operations on Bitcoin.

  \subsection{Transaction fees}
\label{sec:api:fees}

\begin{figure}[t]
  \centering
  \resizebox{\columnwidth}{!}{%
    \begin{minipage}{1.1\columnwidth}
      \inputBDLlinenos{sources/fees_BTC_MongoDB.txt}
    \end{minipage} 
  }
  \vspace{-10pt}
  \caption{Exposing transaction fees.}
  \label{lst:api:fees}
\end{figure}

In this~\namecref{sec:api:fees} we study \emph{transaction fees},
which are earned by miners when they append a new block to the blockchain.
Each transaction in the block pays a fee,
which in Bitcoin is
defined as the difference between its input and output values.
While the values of outputs are stored explicitly in the transaction,
those of inputs are not:
to obtain them, one must retrieve from a past block the transaction that is redeemed by the input.
This can be obtained through a ``deep'' scan of the blockchain, 
which is featured by our library.
We show in~\Cref{lst:api:fees} how to construct a collection 
which contains, for each transaction:
\begin{inlinelist}
\item the hash of the enclosing block;
\item the transaction hash;
\item the fee;
\item the date in which the transaction was appended to the blockchain;
\item the exchange rate between \BTC and \USD in such date.
\end{inlinelist}

The extra parameter \code{true} in the \code{BitcoinSettings} constructor
(missing in the previous example), triggers the ``deep'' scan. 
When scanning the blockchain in this way, 
the library maintains a map which associates transaction outputs to their values,
and inspects this map to obtain the value of inputs%
\footnote{Since inputs can only redeemed transactions on past blocks, the map always contains the required output.
  Although coinbase inputs do not have a value in the map, 
  we calculate their value using the total fees of the current block and the block height 
  (reward is halved each 210,000 blocks).}.
The methods \code{getInputsSum} (resp., \code{getOutputsSum}) at line~\lineno{10}
returns the sum of the values of the inputs (resp., the outputs) of a transaction.

The obtained collection can be used to perform several standard statistics,
\eg the daily total \href{https://blockchain.info/charts/transaction-fees}{transaction fees},
the \href{https://bitinfocharts.com/comparison/bitcoin-transactionfees.html}{average fee},
the percentage earned by miners from transaction fees, \etc
Here we analyse the so-called
\emph{whale transactions}~\cite{Liao17incentivizing}, 
which pay a unusually high fee to miners.
To obtain the whale transactions, 
we first compute the average $\bar{x}$ and standard deviation $\sigma$ 
of the fees in all transactions: 
in \USD, we have $\bar{x} = 0.41$, $\sigma = 12.09$.
Then, we define whale transactions as those which
pay a fee greater than $\bar{x}+2\sigma=24.58$ \USD. 
Overall we collect $242,839$ whale transactions; 
those with biggest fee are displayed in~\Cref{fig:whale}.

\begin{figure}[h]
  \begin{center}
    \begin{minipage}[h]{\columnwidth}
      \centering
      \resizebox{\columnwidth}{!}{
        \begin{filecontents*}{queries/whaleTransactions_BTC_MongoDB_Result1.csv}
Fee (USD), Date, Transaction hash
136243.37, 2016-04-26 14:15:22, cc455ae816e6cdafdb58d54e35d4f46d860047458eacf1c7405dc634631c570d
56493.50, 2017-01-04 20:01:28, d38bd67153d774a7dab80a055cb52571aa85f6cac8f35f936c4349ca308e6380
39502.15, 2017-05-31 14:28:51, cb95ab3aef378c14bc59d0db682d96202b981c1f8fad7d66e23e0be06f2a00c4
25095.71, 2017-05-31 14:28:51, 8e12a1aba87e4657f5fabec1121ed57f706805ad6d4ffe88c6fce78596bd9b75
23518.00, 2013-08-28 10:45:17, 4ed20e0768124bc67dc684d57941be1482ccdaa45dadb64be12afba8c8554537
\end{filecontents*}
\csvautotabular{queries/whaleTransactions_BTC_MongoDB_Result1.csv}
      }
    \end{minipage}
  \end{center}
  \vspace{-10pt}
  \caption{The five biggest whale transactions.}
  \label{fig:whale}
\end{figure}

  \subsection{Address tags}
\label{sec:api:tags}

\begin{figure}[t]
  \centering
  \resizebox{\columnwidth}{!}{%
    \begin{minipage}{\columnwidth}
      \inputBDLlinenos{sources/tags_BTC_SQL.txt}
    \end{minipage}
  }
  \vspace{-10pt}
  \caption{Associating transaction outputs with tags (SQL version).}
  \label{lst:api:tags}
  \vspace{-0pt}
\end{figure}

The webpage \href{https://blockchain.info/tags}{\code{blockchain.info/tags}}
hosts a list of associations between Bitcoin addresses and \emph{tags}
which briefly describe their usage%
\footnote{For instance,	
  address \href{https://blockchain.info/address/1PQCrkzWweCw4huVLcDXttAZbSrrLbJ92L}{1PQCrkzWweCw4huVLcDXttAZbSrrLbJ92L}
  is associated to tag \textit{Linux Mint Donations}
  \url{http://www.linuxmint.com/donors.php}
}.
\Cref{tab:related} shows that address tags are widely adopted,
\eg analytics for cyber-crime usually retrieve addresses tagged as scam or ransomware on forums;
market analyses exploit tags for recognising addresses of business services;
anonymity studies tag the addresses that seem to belong to the same entity.
In this~\namecref{sec:api:tags} we construct a blockchain view 
where outputs are associated with the tags of the address which can redeem them
(we discard the outputs with untagged addresses).
More specifically, we construct an SQL table 
whose columns represent transaction outputs containing: 
\begin{inlinelist}
\item hash of the enclosing transaction;
\item the date in which the transaction has been appended to the blockchain;
\item the output value (in \BTC);
\item the address receiving the payment;
\item the tag associated to the address.
\end{inlinelist}

\Cref{lst:api:tags} shows the Scala script which builds this table. 
At line~\lineno{1}, we connect to the MySQL database.
We retrieve tags from an external source, the \code{blockchain.info} website.
While in the previous case studies we have retrieved external data
by querying the source (\eg the Coindesk APIs),
in this case we query a local file in which 
we have stored the data fetched from~\code{blockchain.info}.
At line~\lineno{2}, given the file containing tags, 
the \code{Tag} class builds a \code{Map} 
which associate each address to the correspondent tag.
At lines~\lineno{4-11} we create a new table.
At lines~\lineno{13-15} we iterate over all the transaction outputs.
At line~\lineno{16} we try to extract the address 
which can redeem the current output.
If we find it (line~\lineno{17}), 
then we search the map for the associated tag (line~\lineno{18});
if a tag is found (line~\lineno{19}) 
we insert a new row into the \code{tagsoutputs} table (line~\lineno{20}).

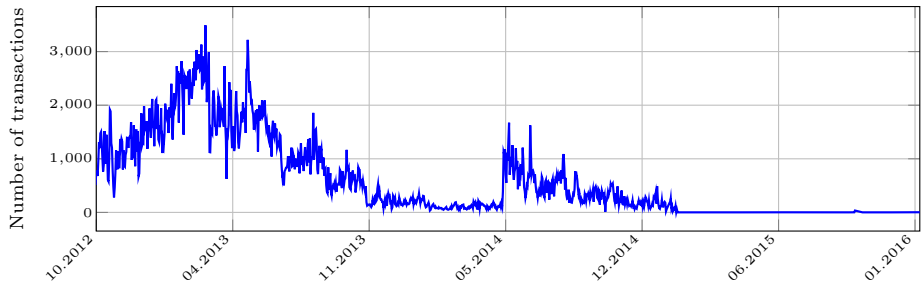
\begin{figure}[h]
	\centering
	\scriptsize
	\resizebox{\columnwidth}{!}{%
		\begin{tikzpicture}
		\begin{axis}[
		width  = 1\linewidth,
		height = 4.5cm,
		date coordinates in=x, date ZERO=2012-10-1,
		xmin=2012-10-1, xmax=2016-1-21,
		xmajorgrids = true,
		ymajorgrids = true,
		xticklabel style={rotate=45, anchor=east, xticklabel}, xticklabel=\tiny\month.\year,
		yticklabel style={font=\tiny,xshift=0.5ex},
		ylabel absolute, ylabel style={yshift=-0.2cm}, ylabel={\scriptsize Number of transactions}
		]
		\pgfplotstableread[col sep=comma]{queries/tags_BTC_MongoDB_Result1.csv}\data
		\addplot [color=blue, thick] table[x index = {0}, y index = {1}]{\data};
		\end{axis}
		\end{tikzpicture}
	}
	\vspace{-15pt}
	\caption{Number of daily transactions to addresses tagged with \textit{SatoshiDICE}*.}
	\label{fig:pgf:tags}
\end{figure}

Using the obtained view, one can aggregate transactions on different 
business levels~\cite{lischke2016analyzing} 
to obtain statistics about the total number of transactions,
the amount of \BTC exchanged, 
the geographical distributions of tagged service, \etc
In particular, we aggregate all addresses whose 
tag starts with \textit{SatoshiDICE}, 
and then we measure the number of daily transactions which send \BTC to 
one of these addresses.
The diagram in~\Cref{fig:pgf:tags} shows the results of this analysis.
The fall in the number of transactions at the start of 2015 may be due
to the fact that SatoshiDICE is using untagged addresses.

\section{Implementation and validation}
\label{sec:evaluation}

We implement the Ethereum-side of our library by exploiting Parity,
queried by means of the \href{https://github.com/web3j/web3j}{web3j library}.
Bitcoin data is provided by both BitcoinJ and the RPC interface of Bitcoin Core.
While BitcoinJ APIs only allow the programmer to retrieve a block by its hash, 
Bitcoin Core's interface exposes calls to do so by its height on the chain.
Furthermore, BitcoinJ's ``block objects'' do not carry information about block height
and the hash of the next block 
(they only have backward pointers, as defined in the blockchain),
which can be fetched by using Bitcoin Core.
Our APIs allow to navigate blockchains.
Particularly, in the Bitcoin case, we do this by iterating over these steps: 
\begin{inlinelist}
	\item get the hash $h$ of the block of height $i$, by using Bitcoin Core;
	\item get the block with hash $h$, by using BitcoinJ;
	\item increment $i$.
\end{inlinelist}
By default, the loop starts from $0$ and stops at the last block.
The methods \code{blockchain.start(i)}, and \code{blockchain.end(j)}
allow to scan an interval of blockchains,
as shown in \Cref{sec:api:opreturn}.
We write the SQL queries exploiting \href{http://scalikejdbc.org/}{ScalikeJDBC},
a SQL-based DB access library for Scala. 
ScalikeJDBC provides also a DSL for writing SQL queries.

We carry out our experiments using consumer hardware,
\ie a PC with a quad-core Intel Core i5-4440 CPU @ 3.10GHz,
equipped with 32GB of RAM and 2TB of hard disk storage.
All the experiments scan the Bitcoin blockchain from the origin 
block up to block number \mylastblock~(added on \mylastdate).
\Cref{tab:evaluation} displays a comparison 
of the size of each view,
and the time required to create and query it.

\begin{table}[t]
	\centering
		\begin{tabular}{|c|c|c|c|c|c|c|}
			\hline
			\multirow{2}{*}{Case study} & \multicolumn{3}{c|}{MongoDB} & \multicolumn{3}{c|}{MySQL} \\ \cline{2-7}
			                            & Create & Query  &    Size    & Create & Query &   Size    \\ \hline\hline
			        Basic view          &  9 h   & 2860 s &   300 GB   &  9 h   & 3.5 h &  266 GB   \\
			    \opreturn metadata      &  2 h   & 0.5 s  &   0.5 GB   & 1.4 h  & 2.5 s &  0.5 GB   \\
			      Exchange rates        &  5 h   & 477 s  &   34 GB    & 4.5 h  & 243 s  &   27 GB   \\
			     Transaction fees       &  9 h   & 448 s  &   51 GB    & 8.5 h  & 614 s &  43.5 GB  \\
			       Address tags         &  4 h   & 1.8 s  &   0.8 GB   & 2.3 h  & 2.7 s &  0.6 GB   \\ \hline
		\end{tabular}%
	\caption{Performance evaluation of our framework.}
	\label{tab:evaluation}
	\vspace{-10pt}
\end{table}

Note that the size of the blockchain view constructed in~
\code{Basic} (\Cref{sec:api:myblockchain})
is more than twice than the current Bitcoin blockchain.
This is because, while Bitcoin stores scripts in binary format, 
our library writes them as strings, so to allow for constructing 
indices and performing queries on scripts.
Moreover, the SQL query in \code{Basic} is particularly 
slow because of the join operations it performs.
Note instead that the query times in SQL and MongoDB are quite similar in all the other cases,
where no join operation is required.

\section{Comparison with related tools}
\label{sec:comparison}

We now compare other general-purpose blockchain analysis tools with ours.
\Cref{tab:comparison} summarises the comparison,
focussing on the target blockchain, the DBMS used, 
the support for creating a custom schema,
and for embedding external data.
The rightmost column indicates the date of the most recent commit in the repository.
Note that all the tools which support Bitcoin
also work on Bitcoin-based \emph{altcoins}.

The projects 
\code{\href{https://github.com/znort987/blockparser}{blockparser}}
and
\code{\href{https://github.com/mikispag/rusty-blockparser}{rusty-blockparser}}
allow one to perform full scans of the blockchain, 
and to define custom listeners which are called each time 
a new block or transaction is read.
Unlike our library, these tools offer limited built-in support
for combining blockchain and external data.
The website \code{blockchainsql.io} has a GUI through which 
one can write and execute SQL queries on the Bitcoin blockchain.
This is the only tool, among those mentioned in~\Cref{tab:comparison},
that does not need to store a local copy of the blockchain.
A drawback is that the database schema is fixed, hence it is not possible
to use it for analytics which require external data. 
While the other tools store results on secondary memory, 
\code{blockparser} and \code{BlockSci} keep all the data in RAM.
Although this speeds up the execution, 
it demands for ``big memory servers'',
since the size of the blockchains of both Bitcoin and Ethereum 
has largely surpassed the amount of RAM available on consumer hardware.
Note instead that the disk-based tools also work on consumer hardware.
Some low-level optimizations, combined with an in-memory DBMS,
help~\cite{Kalodner17arxiv} to overwhelm the performance of the disk-based tools. 
Unlike the other tools, \cite{Kalodner17arxiv} 
provides also data about transactions broadcast on the peer-to-peer network.

\begin{table}[t]
  \centering
  \resizebox{\columnwidth}{!}{%
    \begin{tabular}{|c|c|c|c|c|c|}
    	\hline
    	    \textbf{Tool}      & \textbf{Blockchain} &    \textbf{Database}    & \textbf{Schema} & \textbf{Ext.\ data} & \textbf{Updated} \\ \hline\hline
    	     blockparser       &         BTC         &        RAM-only         &     Custom      &       Custom        &     2015-12      \\
    	  rusty-blockparser    &         BTC         &        SQL, CSV         &      Fixed      &       Custom        &     2017-09      \\
    	   blockchainsql.io    &         BTC         &           SQL           &      Fixed      &        None         &       N/A        \\
    	       BlockSci        &         BTC         &        RAM-only         &     Custom      &       Custom        &     2017-09      \\
    	    python-parser      &         BTC         &          None           &      None       &       Custom        &     2017-05      \\
    	\textbf{Our framework} &  \textbf{BTC, ETH}  & \textbf{MySQL, MongoDB} & \textbf{Custom} &   \textbf{Custom}   & \textbf{2017-09} \\ \hline
    \end{tabular}%
  } %
  \caption{General-purpose blockchain analytics frameworks.}
  \label{tab:comparison}
  \vspace{-10pt}
\end{table}

Remarkably, as far as we know none of the analyses mentioned 
in~\Cref{tab:related} uses the general-purpose tools in~\Cref{tab:comparison}.
Instead, several of them acquire blockchain raw data 
by using Bitcoin Core%
\footnote{\url{https://bitcoin.org/en/bitcoin-core}. 
  Another popular tool for accessing the blockchain was 
  Bitcointools (\url{https://github.com/gavinandresen/bitcointools}), 
  but it seems no longer available.}
(the reference Bitcoin client),
and encapsulate them into Java objects with
the \href{https://bitcoinj.github.io}{BitcoinJ APIs} before processing.
However, neither Bitcoin Core nor BitcoinJ are natural tools 
to analyse the blockchain:
the intended use of BitcoinJ is to support the development of
wallets, and so it only gives direct access to blocks and transactions
from their \emph{hash},
but it does not allow to perform forward scans of the blockchain.
On the other hand, Bitcoin Core would provide the means to scan the blockchain,
but this requires expertise on its low-level RPC interface,
and even doing so would result in raw pieces of JSON data, 
without any abstraction layer.

A precise comparison of the performance of these tools against ours
is beyond the goals of this paper.
The performance analysis in~\Cref{tab:evaluation}
is a first step towards the definition of a suite of
benchmarks for evaluating blockchain parsers.

\section{Conclusions and future work}
\label{sec:conclusions}

We have presented a framework for developing general-purpose analytics 
on the blockchains of Bitcoin and Ethereum.
Its main component is a Scala library which can be used to construct 
views of the blockchain, possibly integrating blockchain data
with data retrieved from external sources.
Blockchain views can be stored as SQL or NoSQL databases, 
and can be analysed by using their query languages. 
Our experiments confirmed the effectiveness and generality of our approach, which uniformly comprises in a single framework several use cases addressed by various ad-hoc approaches in literature. 
Indeed, the expressiveness of our framework overcomes that 
of the closer proposals
in the built-in support for external data, 
and the support of different kinds of databases and blockchains. 
Importantly, coming in the form of an open source library for a mainstream language, 
our framework is amenable of being validated and extended by a community effort, following reuse best practices.

On the comparison of SQL vs NoSQL, 
our experiments did not highlight significant differences 
in the complexity of writing and executing queries in the two languages. 
Instead, we observed that the schema-less nature of NoSQL databases 
simplifies the Scala scripts. 
From~\Cref{tab:evaluation} we see that 
both creation and query time are comparable as order of magnitude. 
As already discussed in~\Cref{sec:evaluation},
the difference in the execution time of queries is
due to join operations in SQL.
A more accurate analysis, carried over a larger benchmark, 
is scope for future work.
Anyway, it is worth recalling that the goal of our proposal is provide to the final user the flexibility to choose the preferred database, rather than ascertain an idea of best-fit-for-all in the choice.

Although our framework is general enough to cover most of the
analyses in~\Cref{tab:related}, it has some limitations
that can be overcome with future extensions.
In particular, some analyses
addressing \eg information propagation, forks and attacks~\cite{Decker13p2p,Donet14bitcoin,McCorry16fc,Pappalardo17arxiv}
require to gather data from the underlying peer-to-peer network.
To support this kind of analyses one has to run a customized node
(either of Bitcoin or Ethereum).
Such an extension would also be helpful to
obtain on-the-fly updates of the analyses.
  \paragraph{Acknowledgments.}

This work is partially supported by
Aut.\ Reg.\ of Sardinia P.I.A.\ 2013 ``NOMAD''.
This paper is based upon work from COST Action IC1406 cHiPSET, 
supported by COST (European Cooperation in Science and Technology).

\bibliographystyle{splncs03}
\bibliography{main}

\end{document}